\documentclass[journal]{IEEEtran}
\usepackage{url}
\usepackage[utf8]{inputenc}
\usepackage{xcolor}
\usepackage{amsmath}
\usepackage{amssymb}

\usepackage[acronyms,nonumberlist,nopostdot,nomain,nogroupskip]{glossaries}
\usepackage{tablefootnote}
\usepackage{booktabs}
\usepackage{tabularx}
\usepackage{tikz}
\usepackage{pgfplots}
\pgfplotsset{compat=newest} 
\pgfplotsset{plot coordinates/math parser=false} 
\newlength\fheight
\newlength\fwidth
\usetikzlibrary{plotmarks,patterns,decorations.pathreplacing,backgrounds,calc,arrows,arrows.meta,spy,matrix}
\usepgfplotslibrary{patchplots,groupplots}
\usepackage{tikzscale}
\usepackage{hyperref}

\usepackage{multirow}
\usepackage[font=scriptsize]{subcaption}
\usepackage[font=footnotesize]{caption}

\usepackage{mathtools}

\usepackage{dblfloatfix}    
\usepackage{colortbl}

\usepackage{makecell}
\usepackage{diagbox}
\usepackage{tikz-qtree}
\usetikzlibrary{trees} 
\usepackage{cite}
\usepackage{bbold}
\usepackage{bbm}

\usepackage{array}
\newcolumntype{?}{!{\vrule width 1.5pt}}
\newcolumntype{P}[1]{>{\centering\arraybackslash}p{#1}}
\usepackage{makecell}
\usepackage{mdframed}
\usepackage[many]{tcolorbox}
\usepackage{enumitem}
\usepackage{booktabs}

\newtcbox{\mybox}[1][]{nobeforeafter,math upper,tcbox raise base,
  enhanced,frame hidden,boxrule=0pt,interior style={top color=green!10!white,
  bottom color=green!10!white,middle color=green!50!yellow},
  fuzzy halo=1pt with green,drop large lifted shadow,#1}

\usepackage{siunitx}

\usetikzlibrary{fadings}

\tikzfading[name=middle,
            top color=transparent!100,
            bottom color=transparent!100,
            middle color=transparent!20]

\usetikzlibrary{arrows,automata,calc,shapes, positioning,shadows,shadows.blur,shapes.geometric}

\makeglossaries

\makeatletter
 \let\old@ps@headings\ps@headings
 \let\old@ps@IEEEtitlepagestyle\ps@IEEEtitlepagestyle
 \def\confheader#1{%

 \def\ps@IEEEtitlepagestyle{%
 \old@ps@IEEEtitlepagestyle%
 \def\@oddhead{\strut\hfill#1\hfill\strut}%
 \def\@evenhead{\strut\hfill#1\hfill\strut}%
 }%
 \ps@headings%
 }
 \makeatother

\confheader{%
\begin{minipage}{\textwidth}
\centering \scriptsize
This paper has been accepted for publication in IEEE Communications Standards Magazine, ©2022 IEEE.\\
A. Chaoub, A. Mämmelä, P. Martinez-Julia, R. Chaparadza, M. Elkotob, L. Ong, D. Krishnaswamy, A. Anttonen, A. Dutta, “Hybrid Self-Organizing Networks: Evolution, Standardization Trends, and a 6G Architecture Vision”, IEEE Communications Standards Magazine, 2022.
\end{minipage}
 }
 
\begin{document}
\pagenumbering{gobble}


\title{Hybrid Self-Organizing Networks: Evolution, Standardization Trends, and a 6G Architecture Vision}

\author{{{Abdelaali Chaoub},~\IEEEmembership{Senior Member, IEEE},
{Aarne Mämmelä},~\IEEEmembership{Senior Member, IEEE},\\
{Pedro Martinez-Julia},~\IEEEmembership{Member, IEEE},
{Ranganai Chaparadza},
{Muslim Elkotob},~\IEEEmembership{Member, IEEE},\\
{Lyndon Ong},~\IEEEmembership{Member, IEEE}, {Dilip Krishnaswamy},~\IEEEmembership{Senior Member, IEEE}, {Antti Anttonen},~\IEEEmembership{Senior Member, IEEE},\\ {Ashutosh Dutta},~\IEEEmembership{Fellow, IEEE}
}

\thanks{
Abdelaali Chaoub is with the National Institute of Posts and Telecommunications (INPT), Morocco (email: chaoub.abdelaali@gmail.com).
Aarne Mämmelä and Antti Anttonen are with the VTT Technical Research Centre of Finland, Finland (email: firstname.lastname@vtt.fi).
Pedro Martinez-Julia is with the National Institute of Information and Communications Technology (NICT), Japan (email: pedro@nict.go.jp).
Ranganai Chaparadza is with Capgemini Engineering \& IPv6 Forum, Germany (email: ran4chap@yahoo.com).
Muslim Elkotob is with Vodafone, Germany (email: muslim.elkotob@vodafone.com).
Lyndon Ong is with Ciena, USA (email: lyong@ciena.com).
Dilip Krishnaswamy is with Sterlite Tech, USA (email: dilip@ieee.org).
Ashutosh Dutta is with Johns Hopkins University Applied Physics Laboratory, USA (email: ashutosh.dutta@ieee.org)
}
}

\maketitle

\begin{abstract}
Self-organizing networks (SONs) need to be endowed with self-coordination capabilities to manage the complex relations between their internal components and to avoid their destructive interactions. Existing communication technologies commonly implement responsive self-coordination mechanisms that can be very slow in dynamic situations. The sixth generation (6G) networks, being in their early stages of research and standardization activities, open new opportunities to opt for a design-driven approach when developing self-coordination capabilities. This can be achieved through the use of hybrid SON designs. A hybrid architecture combines the centralized and distributed management and control. In this article, we review the history of SONs including the inherent self-coordination feature. We then delve into the concept of hybrid SONs (H-SONs), and we summarize the challenges, opportunities, and future trends for H-SON development. We provide a comprehensive collection of standardization activities and recommendations, discussing the key contributions and potential work to continue the evolution and push for a wide adoption of the H-SON paradigm. More importantly, as a key 6G architectural feature we propose that H-SONs should be loosely coupled networks. Loose coupling refers to the weak interaction of different layers and weak interaction between users in the same layer, i.e., the various feedback loops must be almost isolated from each other to improve the stability and to avoid chaotic situations. We finally conclude the paper with the key hints about the future landscape and the key drivers of 6G H-SONs.
\end{abstract}


\section{Introduction} 
\label{sec:introduction}
The forthcoming 6G networks aim to solve the problems with global connectivity and sustainability, leading to mutually conflicting objectives and to the increased use of self-organizing networks (SONs), which are still immature. In fact, future 6G networks intend to democratize uniform access to digitalization among the global population to reduce existing social and economic inequalities. However, this leads to an increased wireless connectivity footprint and a pressing demand for additional energy resources, thus reducing sustainability. At the same time, 6G will offer orders of magnitude performance improvements over 5G in cities. Actions to achieve certain network key performance indicators (KPIs) may, however, disturb each other due to the overlapping network parameters. Handling efficiently these conflicting performance metrics, such as the dilemma of coverage vs. energy efficiency and load distribution vs. interference, is beyond the capabilities of current 5G networks. It is imperative that 6G networks be designed as self-organizing from their early design and specification stages to monitor and resolve potential destructive interactions. 

Self-organization~\cite{Aliu},~\cite{Fourati} is a general term that covers any kind of autonomous restructuring of a system and is the highest and most general in the hierarchy of all technical systems~\cite{Mammela}. SONs will therefore continue to empower future 6G communication systems that must offer certain fundamental properties, i.e., functionality, stability, scalability, performance, dependability, security, cost-effectiveness, and resilience~\cite{Dobson}. The SON architecture is decomposed into a set of smaller functional units referred to as SON functions (SONFs). Examples include Coverage and Capacity Optimization (CCO) and Energy Saving Management (ESM) functions~\cite{Bayazeed}. SONFs may interact either constructively or destructively. Thereby, advanced self-coordination capabilities are needed to ensure a conflict-free operation. In recent 5G-related studies, most self-coordination principles are usually not considered at the design phase of new communication systems adopting either a predictive or a reactive approach \cite{Bayazeed}. From a standardization perspective, the focus has been on the 3rd Generation Partnership Project (3GPP), the International Telecommunications Union-Telecommunication Standardization Sector (ITU-T), and the European Telecommunications Standards Institute (ETSI) activities~\cite{Kafle}, thus overlooking many initiatives driven by other standardization development organizations (SDOs).
\begin{figure*}[t!]
\centering
\includegraphics[width=.6\textwidth]{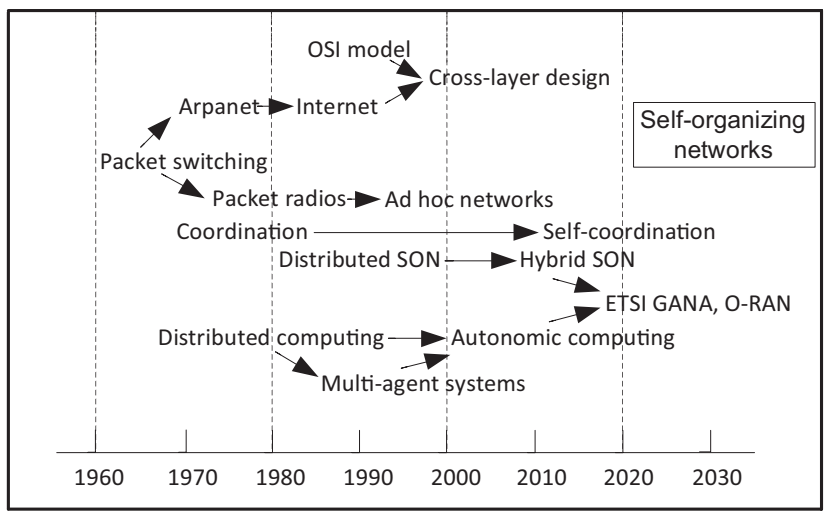}
    \caption{Development of self-organizing networks from packet switching systems to hybrid SONs.}
    \label{fig:history}
\end{figure*}

To the best of our knowledge, the paper proposed by Bayazeed et al.~\cite{Bayazeed} is the first attempt to provide a detailed survey about the self-coordination functionality in cellular networks. It introduces a high-level comprehensive framework to categorize self-coordination logics into protective, reactive, and proactive. According to~\cite{Bayazeed}, protective methods that anticipate conflicts during the design stage are only valid for static situations whereas system dynamics call for proactive methods that can predict the potential conflicts at the execution time using artificial intelligence (AI) and machine learning (ML). 

\begin{figure*}[t!]
\centering
\includegraphics[width=.9\textwidth]{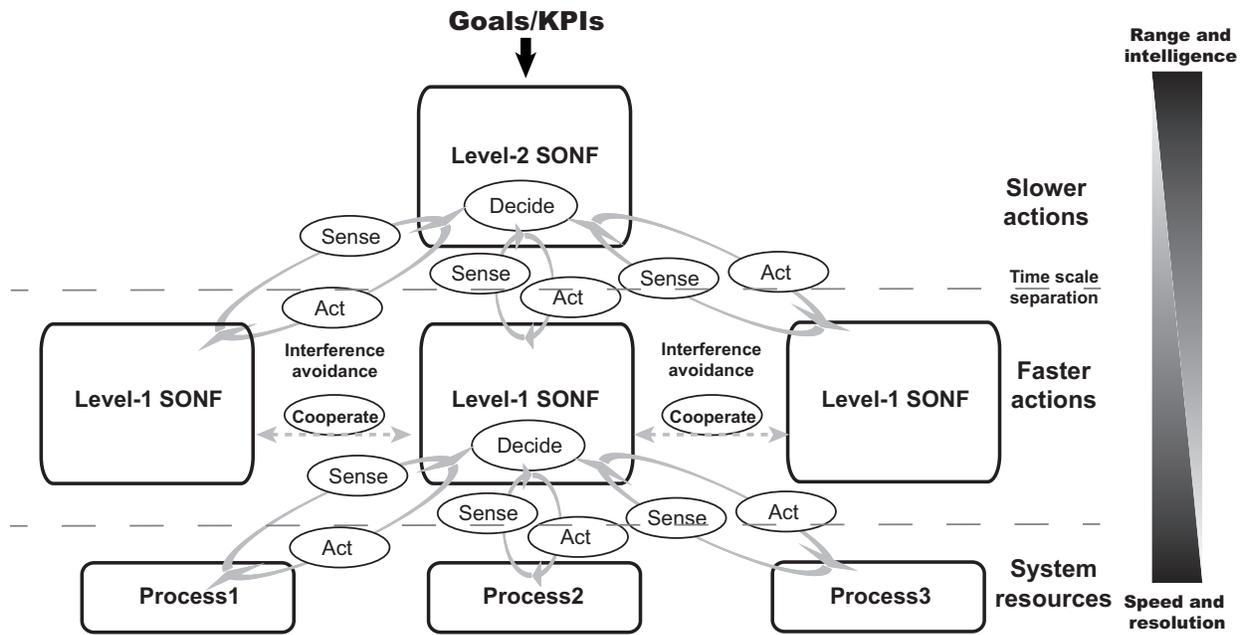}
    \caption{Improved hierarchy using loosely coupled system design.}
    \label{fig:H-SON}
\end{figure*}

Along these lines, evolved loosely coupled hybrid SONs (H-SONs) are a key concept that is used throughout this paper. The H-SON design \cite{Fourati}, \cite{Bayazeed} has been introduced by 3GPP in release 8 as a mix of both centralized and distributed SON architectures, whereas loose coupling \cite{mammela2022new} refers to the weak interaction of different layers and weak interaction between users in the same layer. Decoupling of feedback loops has been known in control theory since the 1960s, but it is a rather new idea in the higher layers of communication systems as can be seen in the design of Open Radio Access Network (O-RAN) using cross-layer optimization. In conventional networks, the feedback loops have worked strictly within each layer and even there instability may be caused by large delays or horizontal coupling. We argue that future SONs should adaptively choose the right tradeoff of centralized and distributed management and control while maintaining a loose coupling between the underlying horizontal and vertical system components. This promising architectural model will achieve a stable and almost conflict-free SON operation by design as opposed to reactive and proactive methods that can be too slow in dynamic situations. We then elaborate relations to the recent O-RAN concept, a promising application framework for the H-SON design. Such future loosely coupled H-SONs need to be supported by the highly widespread virtualization, containerization, and multi-tenant architectures, but it must be further empowered by the incorporation of quantum technology, contextual awareness and the latest advancements in AI. Finally, the technical definitions in future H-SONs will also address energy efficiency and general SDOs by enforcing optimality and policy-based evolution, in which humans play a key role.

Accordingly, the contributions of this article are two-fold. First, we outline the architectural innovations required as well as the associated drivers suited for 6G networks to achieve an improved design-driven self-coordination. Second, we provide a comprehensive summary of standardization initiatives related to SONs and their underlying self-coordination aspects to highlight the interest in a harmonized framework. Broader standardization-related insights are presented in~\cite{IEEE_SysOpt_WG}.

The remainder of this paper is organized as follows. In Section~\ref{sec:Background_and_State_of_the_art}, we present the background and state-of-the art of SONs. Then, Section~\ref{sec:Opportunities_Use_Cases_Technical_Hurdles_and_Standardization_Landscape} discusses the opportunities, use cases, and standardization landscape of SONs. The challenges, trends as well as directions of research and standardization activities regarding SONs are identified in Section~\ref{sec:Challenges_Trends_and_Research_Standard_Directions}. Finally, Section~\ref{sec:conclusions} concludes the paper.

\section{Background and State-of-the-art} 
\label{sec:Background_and_State_of_the_art}
The historical development of SONs is presented in Fig.~\ref{fig:history}. To obtain intelligent behavior, a feedback loop is needed. It is based on the sense-decide-act loop, which in modern AI forms the basis of an intelligent and rational agent, for example an SONF. The idea led to multi-agent systems, which form a hierarchy of interacting agents, a good basis for self-organization. In a hierarchy, the basic idea is to use low speed in the higher layers such as in the network layer and high speed in the lower layers such as in the physical layer. The speed should correspond to the rate of change of the relevant network parameters in each layer. In a hierarchical system, a form of joint optimization is cross-layer optimization.


In communications, the first SON was the Internet that was originally called the Arpanet and is based on packet switching, a form of loose coupling. At the same time, the generalized Open Systems Interconnection (OSI) model was developed. Ad hoc networks are multi-hop SONs without any fixed infrastructure. The history of ad hoc networks started from packet radio networks. The terms vertical and horizontal loose coupling were explicitly defined by Simon in the beginning of 1970s, but the idea of loose coupling of feedback loops has stayed mostly within control theory.

The interest in SONs rose in the 1980s and were originally defined to be distributed. The 3GPP Rel. 8 (2008) divided SONs into three groups, including centralized SONs (C-SONs), distributed SONs (D-SONs), and H-SONs, which are a combination of C-SON and D-SON. 3GPP Rel. 11 (2011) proposed the term self-coordination to avoid and resolve conflicts in SONs~\cite{Bayazeed}. The research on coordination of conflicts in hierarchical systems started in the 1960s with Mesarovic’s pioneering work.

Independently of SONs, distributed computing was developed in computer science towards autonomic computing that is based on self-management. The term was later adopted also in communication networks. ETSI defined the Generic Autonomic Networking Architecture (GANA) reference model as an ETSI Technical Specification (TS) ~\cite{Etsi_103_195}, which corresponds to the H-SON concept defined earlier by 3GPP.

C-SONs are based on forced cooperation of network nodes using hierarchy. In D-SONs the nodes cooperate at least with their closest neighbors by exchanging information. H-SONs combine centralized and distributed control which means that in addition to having hierarchy, the nodes exchange information with neighboring nodes. In C-SONs, optimality can be attained using the global view, but the control overhead and delays are large, and stability and energy efficiency must be compromized. D-SONs do not have any global view and thus they are only locally optimal, but the control overhead and delays are small, and stability and energy efficiency are improved. H-SONs have same advantages compared to C-SONs but also improved global optimality compared to D-SONs. The main disadvantage of H-SONs comes from the flexibility that reduces the global energy efficiency compared to D-SONs, but balances the use of energy consumption between the nodes.

A schematic view of loosely coupled H-SON architecture is provided in Fig.~\ref{fig:H-SON}, showing essentially a hierarchical loosely controlled set of almost autonomous agents. There is only a loose vertical and horizontal coupling or interaction between agents to improve stability so that the feedback loops are loosely coupled or almost isolated from each other \cite{mammela2022new}. Using different degrees of coupling, the hybrid form can implement all other degrees of centralization as special cases. SONs should have at least a loose centralized control to improve optimality, fairness, and stability so that the global behavior is predictable. Accordingly, the higher in the hierarchy we are, the more intelligence and complexity we have.

Coupling may result in stability problems that can be avoided by keeping the degree of coupling, denoted here by $C$, small enough. It can be quantitatively measured. In horizontal coupling it can be defined as the ratio of leaked interfering power and total received power after possible filtering, which corresponds to the inverse of signal-to-interference ratio in communication systems. In vertical coupling, the degree of coupling can be defined as the ratio of speed of the higher layer and the speed of the next lower layer. The speed can be measured using the bandwidth of the corresponding changes.

An uncoupled system corresponds to $C = 0$ and fully coupled corresponds to $C = 1$. A loosely coupled system corresponds to $0 < C << 1$ and a tightly coupled system corresponds to $0 << C < 1$. In general, the requested numerical value for $C$ depends on performance requirements. A starting point could be $C = 0.001$ for loose coupling but it could be for example $C = 0.01$ with more interference and reduced stability. Small $C$ can be obtained by time-scale separation between layers and interference avoidance in the same layer, usually physical layer. These initial definitions form the basis for a complete framework that we leave for future studies.

The degree of centralization of the 3GPP cellular networks has been evolving significantly with the change of different generations from 2G to 5G. The optimal balancing between fully centralized and decentralized network architecture is important because it has high influence to the overall system performance. The degree of centralization concerns three separate network domains, namely radio access network (RAN), packet core network (PCN), and cloud computing network (CCN). In 2G networks, all radio frequency (RF) and baseband (BB) parts were implemented in the base stations (BSs) apart from antennas. In 3G, more distributed RAN started to appear as the RF and BB parts were separated by moving RF parts closer to the antenna to reduce the RF losses. With 4G, the PCN architecture took a step towards a more distributed direction, as BSs were enabled to communicate directly without PCN via the X2 interface. At the same time, many latency-critical network control functions were distributed to BSs. In 5G, user data and network control planes were separated. Moreover, the flexibility to distribute PCN functionalities was increased by allowing more control functions to be in separate distributed control units closer to users. On the other hand, the RAN continues to move to a more centralized mode by adopting centralized and virtualized resource pools to manage resources more efficiently. The previously centralized CCN was extended into a more distributed mobile edge computing platform. The key enablers towards these directions are the logically centralized software-defined networking (SDN) and virtual network function (VNF) concepts. Finally, in the upcoming 6G network, the aim is a deeper integration of non-terrestrial aerial and satellite networks in terms of network management~\cite{Darwish}. Therein, the location optimization of network controllers will become even more important due to larger distances and delays between network nodes, potentially increasing both control delay and overheads if not properly designed. In summary, there is no clear direction for the degree of centralization, but the evolution is driven by the challenge on finding optimal hybrid solutions for all three network domains. A typical 5G node is estimated to have 2000 configuration parameters~\cite{Bayazeed}.

Quality of Service (QoS) metrics and targets are seriously compromised by SONF conflicts, and are not well covered by current management and self-* approaches in 5G. For instance, both 5G and 6G network services need precise assessment of delay, jitter, packet loss rate, and communication capacity. Furthermore, additional Quality of Experience (QoE) metrics are particularly required for 6G, such as, level of user satisfaction and response time from users. The integration of non-network data such as location-based (i.e., analysis of location and behavior of users) and social-based (i.e., analysis of social data) analytics into the SON stack, and continuously monitoring the key quality indicators (KQIs) \cite{8868102} of the provisioned service will improve user satisfaction. However, the accurate estimation of the KQIs of different services is still hampered by the increasingly stringent delay requirements of feedback loops. The H-SON model using loose coupling is a significant step towards addressing these challenges through decoupling network optimization-related SONFs from their QoS/QoE counterpart.

\section{Opportunities, Use Cases, and Standardization Landscape} 
\label{sec:Opportunities_Use_Cases_Technical_Hurdles_and_Standardization_Landscape}
In this section we discuss some opportunities that may bring new possibilities, emerging use cases along with  the standardization issues and initiatives derived from them.
\smallskip

\subsection{Opportunities}
Some SONFs mostly focus on short term optimization as they mainly operate in the lower layers. With the aid of pervasive intelligence, SONFs will be able to process both historical information gathered from previous experiences and surrounding contextual data to build solid long-term optimization and prediction, and thus will truly operate the network in a proactive and autonomous fashion.

On the other hand, a key opportunity is to work on the combined hierarchy and degrees of centralization to improve self-coordination. Hierarchy is used in SONs to manage the inherent complexity whereas different degrees of centralization maintain SONs scalability. H-SONs (Fig.~\ref{fig:H-SON}) are envisioned to avoid the weaknesses of pure centralized and distributed paradigms. Such hierarchical networks can be created statically in a structured manner or dynamically in an unstructured manner when needed. Structured network partitions can provide better support for AI/ML within and across network partitions. 

The intrinsic priorities of hierarchical systems are a very powerful mechanism to resolve conflicts in SONs. Sensing information propagates upwards and control information downwards in the hierarchy so that higher layers will detect the conflict and act by preempting the lower layers where decisions are usually made because they are faster. This hierarchy will prevent a conflict from lasting for a long time causing system performance degradation. However, hierarchical systems may need a delay to detect, analyze, and trigger the best countermeasure commensurate with the number of layers.
\smallskip
\begin{figure*}[t!]
\centering
\includegraphics[width=.7\textwidth]{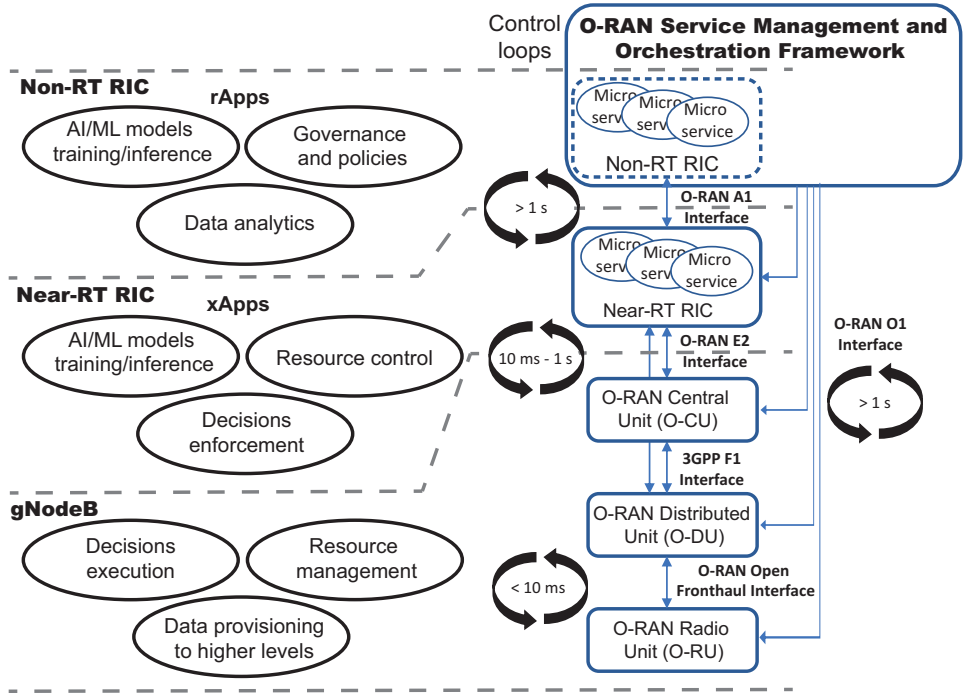}
    \caption{Open RAN use case control loops hierarchy.}
    \label{fig:O-RAN}
\end{figure*}

\subsection{Use Cases} 
The O-RAN architecture has been deliberately designed with flexibility to deploy SON functions at the core, metro or edge depending on the control loop latency requirements, using the non real-time (Non-RT) RAN Interface Controller (RIC), near real-time (Near-RT) RIC and potential use of edge platforms hosting VNFs as hosts for AI/ML applications as well (Fig.~\ref{fig:O-RAN}).

Development of third-party SON software realized as rApps and xApps will be made easier by the definition of reference designs for the supporting platforms' hardware (e.g., commercial off-the-shelf (COTS) equipment) and software and through the definition of interface standards for the sensing input and action output of SONFs in different layers (e.g., O1 and E2 interfaces).

O-RAN design envisions the use of the H-SON concept where there is interaction between SONFs at different positions in the network. The definition of the A1 interface, in particular, allows a higher-level SON in the Non-RT RIC to access information from the Near-RT RIC and to exert policy controls over the Near-RT RIC. Support of integrated SONFs has been adopted as one of the use cases of interest within O-RAN. The key high-level O-RAN functions are depicted in Fig.~\ref{fig:O-RAN} allowing for a multitude of more specialized use cases ranging from RAN sharing to unmanned aerial vehicles (UAVs) radio resource allocation using a loosely coupled design.
\smallskip

\subsection{Standardization Landscape}
SONs are a key target of SDOs~\cite{Kafle},~\cite{Long}. They are included in the key framework of the ITU-T (Rec. ITU-T Y.3324) as a target for Long Term Evolution (LTE) and beyond, thus SON becomes a source of standardization efforts for 4G successors. As shown in~\cite{Martinez_Julia}, some features of SONs have been envisioned for the overall standardization of networks that self-adapt to changes in their environment, being aligned with ETSI Network Function Virtualization (NFV) Management \& Orchestration (MANO) and assumed by the Internet Engineering Task Force (IETF) as part of the Network Management Research Group (NMRG) and the Autonomic Networking Integrated Model and Approach (ANIMA) working group (WG). Standardizing self-organization and intelligent reasoning procedures are essential, so all components (hardware and software) must be assessed in terms of their SON capabilities (i.e., auto-*, self-*), as well as their overall adequacy and quality, for the tenant of the network to have a high degree of certainty on the network to operate as expected.

ETSI efforts towards SON is taking part as Zero Touch Network and Service Management (ZSM), Experiential Network Intelligence (ENI)~\cite{Kafle}, and GANA~\cite{Etsi_103_195}. They provide underlying mechanisms to build SONs but there are still many topics to work on, as defined by ETSI in TS 128 313. They focus on optimization and configuration management but there are open issues on interaction among ETSI, ITU-T, and IETF views needed to define a stable framework for SONs~\cite{IEEE_SysOpt_WG}, such as the interface compatibility and adaptation, the function definition for consistency, and the consolidation of architecture components for allowing multi-vendor deployments.


\begin{table*}[t!]
\centering
\footnotesize
\setlength{\belowcaptionskip}{0.33cm}
\caption{Current and future standardization activities.}
\label{tab:std_activities}
\renewcommand{\arraystretch}{1.5}
\begin{tabular}{|m{1cm}|m{4cm}|m{6cm}|m{4cm}|m{1cm}|}
\hline
\textbf{SDO or Forum}  & \textbf{WG or Framework } & \textbf{Standardization Activities} & \textbf{Activities Type and Maturity} & \textbf{Launch}\\
\hline

\multirow{2}{2cm}{IEEE} & NGSON WG (P1903 standards) & Service overlay networks as the main abstraction level for autonomics via embracing context awareness and self-organization capabilities. & Regular track & 2013 \\ \cline{2-5}
& INGR SysOpt WG~\cite{IEEE_SysOpt_WG} & Outlines standardization items and approach for enhancing standards on autonomics in other SDOs/fora. & Plans for standardization & 2019 \\ \cline{2-5}
\hline 

\multirow{2}{2cm}{IETF} & ANIMA (e.g., RFC 8993). & Defines a reduced-scope Autonomic Networking (AN) with progressive introduction of autonomic functions (AFs). No implementation specifications for coordination among AFs. & Partly-mature standardization track, inspired and aligned to ETSI GANA. & 2014 \\ \cline{2-5}
& NTF & Architectural framework for network telemetry. Protocols to gather monitoring data for full visibility. & Extend network management beyond conventional OAM. & 2018 \\ \cline{2-5}
& AINEMA & Architectural framework for integrating AI in network management operations. Algorithms to operate AI, information model to represent AI data and decisions, and protocols to exchange them. & Enable network management operations with AI. & 2022 \\ \cline{2-5}
\hline

\multirow{4}{2cm}{ETSI} & TC NTECH/AFI WG and TC INT/AFI WG (e.g., ETSI TS 103 195-2~\cite{Etsi_103_195} and White Paper 16) & GANA model and its instantiations onto various types of fixed, mobile and wireless networks. Running a 5G PoC to implement some GANA aspects. & Regular Standardization track for technical standards and detailed specifications. & 2009 \\ \cline{2-5}

& ENI ISG~\cite{Kafle} & Defines an AI-based architecture to help external systems improve their environmental awareness and adapt accordingly. Envisions the translation of input data as well as output recommendations/commands. & Pre-standardization track. & 2017 \\ \cline{2-5}

& ZSM ISG~\cite{Kafle} & Reuses existing standards and frameworks into a holistic design to achieve E2E automation in multi-vendor  environments using AI-based data collection and closed-loop control. & Pre-standardization track. & 2017 \\ \cline{2-5}
\hline

\multirow{3}{2cm}{ITU} & \multirow{3}{4cm}{SG13~\cite{Kafle}} & Rec. ITU-T Y.3324: defines the functional and architectural requirements of autonomic management and control (AMC) for IMT-2020 networks. & Regular standardization track. & 2018 \\ \cline{3-5}
& & Rec. ITU-T Y.3177: specifies a high-level architecture of AI-based automation of future networks including IMT-2020. & Regular standardization track. & 2021 \\ \cline{3-5}
& & FG-AN: builds upon existing standards' gaps to standardize autonomous networks. & Pre-standardization track. & 2020 \\ \cline{3-5}
\hline

\multirow{4}{2cm}{3GPP} & Release 8 (e.g., TS 32.500) & Basics of LTE-SON. & Mature regular standard. & 2008 \\ \cline{2-5}

& Release 10 (e.g., TS 32.522) & Self-coordination. & Mature regular standard. & 2011 \\ \cline{2-5}


& Release 16 (e.g., TR 28.861) & Introduction to 5G NR-SON. & Mature regular standard. & 2018 \\ \cline{2-5}


& Release 18 & Enhancement of data collection for 5G NR-SON. & Mature regular standard. & 2022 \\ \cline{2-5}
\hline

NGMN & 5G E2E architecture framework v3.0.8 & Describes a high-level vision of architecture principles and requirements to guide other SDOs/Fora and promote interoperability. Its automation capabilities are based on the ETSI GANA model. & Requirements matured as inputs to standards. & 2019 \\ \cline{2-5}
\hline

ONAP & Jakarta (i.e., tenth) Release & Open source model-driven framework that brings orchestration and automation capabilities to physical and virtual network components. & Regular standardization track. & 2017 \\ \cline{2-5}
\hline

BBF & AIM & Builds on GANA and ITU Rs. to define AFs for access and E2E converged fixed/mobile networks. & Regular standards, mature. & 2018 \\ \cline{2-5}
\hline

TMF & ODA (e.g., IG1167 and IG1177) & Mapping of the ETSI GANA framework to the ODA intelligence management model. & Standardized frameworks, mature. & 2018 \\ \cline{2-5}
\hline

\multicolumn{5}{|m{17.8cm}|}{AINEMA = Artificial Intelligence Framework for Network Management; AFI = Autonomic network engineering for the self-managing Future Internet; AIM = Automated Intelligent Management; ANIMA = Autonomic Networking Integrated Model and Approach; BBF = Broadband Forum; ENI = Experiential Networked Intelligence; FG-AN = Focus Group on Autonomous Networks; IMT-2020 = International Mobile Telecommunications-2020; INGR = International Network Generations Roadmap; INT = Core Network and Interoperability Testing; ISG = Industry Specification Group; NGMN = Next Generation Mobile Networks; NGSON = Next Generation Service Overlay Network; NR = New Radio; NTECH = Network Technologies; NTF = Network Telemetry Framework; ODA = Open Digital Architecture; ONAP = Open Network Automation Platform, PoC = Proof Of Concept; Rec. = Recommendation; RFC = Request for Comment; SG = Study Group; SysOpt = Systems Optimization; TC = Technical Committee; TMF = TeleManagement Forum; TR = Technical Report; WG = Working Group; ZSM = Zero touch network \& Service Management.}\\
\hline

\end{tabular}
\end{table*}


Table~\ref{tab:std_activities} summarizes current and future SDO activities, and provides a list of the used abbreviations. It shows that 3GPP, ETSI, Institute of Electrical and Electronics Engineers (IEEE), and IETF have been working for standardizing SON elements for a long time, achieving important success. These are being extended by latest efforts, such as IEEE's International Network Generations Roadmap (INGR)~\cite{IEEE_SysOpt_WG}, ETSI's ZSM, and ITU's Study Group 13 (SG13) which focus on leveraging network autonomicity by standardizing functions and interfaces that enable intelligent operation of network elements. These are also complemented by IETF's AI Framework for Network Management (AINEMA), which presents an information model for integrating AI with network management, and by 3GPP release 18 bringing further enhancements to data collection within 5G SONs. Other initiatives, such as Broadband Forum (BBF) and TeleManagement Forum (TMF), are aligned with GANA. Specifically, BBF joins effort with 3GPP to align their system architectures with GANA at the same time. By its way, the Next Generation Mobile Networks (NGMN) alliance view on SON is aligned with the evolution of 5G toward 6G. It does not align particularly with GANA or ANIMA, although both have key roles in 6G, so NGMN approach will be compatible. Obviously, most of those ongoing standardization initiatives are either compatible or converge toward the reference GANA model, the latter adheres to the H-SON paradigm guidelines.
\smallskip

\section{Challenges, Trends, and Research/Standard Directions} 
\label{sec:Challenges_Trends_and_Research_Standard_Directions}
In this section, we enumerate the fundamental challenges to be addressed for providing conflict-free 6G SONs, and propose novel approaches along with promising research directions. 
\smallskip  

\subsection{Challenges}
\label{subsec:Challenges}
Future SONs need to overcome five key challenges, as summarized by Table~\ref{tab:challenges}: heterogeneity and complexity, loss of control, backward compatibility, fragmented solutions, and expensive optimizations. 

Future communication systems are becoming increasingly complex as a result of supporting new and heterogeneous ecosystems with strong interdependence and huge information flow. Another challenge is that communication systems may act chaotically in unexpected situations, and tight coupling may propagate instability across system components. This increases the risk of losing control and situation-awareness. Advancing self-organization capabilities will also need to preserve the backward compatibility between existing and future standards. Despite the uptake of new communication systems, some fundamental services will continue to rely on legacy technologies as the case of voice services. It is also noteworthy that self-organization is a process that spans various components (e.g., RAN, PCN, CCN) in the end-to-end mobile architecture. This leads to network fragments that can individually self-organize but a holistic SON approach is missing for the whole network. Finally, the full potential of SONs will only be achieved if they can continuously manage computationally expensive optimizations because future systems must provide quasi-instantaneous responsiveness.

\smallskip


\begin{table*}[t!]
\centering
\footnotesize
\setlength{\belowcaptionskip}{0.33cm}
\caption{Future research directions for evolving H-SONs.}
\label{tab:challenges}
\renewcommand{\arraystretch}{1.5}
\begin{tabular}{|m{3.5cm}|m{3cm}|m{4cm}|m{6cm}|}
\hline
\textbf{Challenge}  & \multicolumn{2}{|m{7cm}|}{\textbf{Potential Solutions}} & \textbf{Open Research Questions} \\
\hline
\multirow{3}{3.5cm}{Growing complexity (due to more system elements, heterogeneous ecosystems, hardware dependability, and increased data volumes).} 
& \multirow{3}{3cm}{Evolved H-SON to loosely control various subsystems with reduced intra-network signaling} & Adoption of virtualized, containerized, microservices-based and multi-tenant architectures & Ensure full competition and logical isolation as well as secure and resilient services (i.e., SONFs).\\ \cline{3-4}
& & Use of COTS equipment, open and portable services for multi-vendor interoperability & Risk of obsolescence and need to proactively manage replacement strategies. \\ \cline{3-4}
& & Federated learning, transfer learning and joint learning and communication & Heterogeneity of devices, subsystems, tasks and data. \\ \cline{3-4}
& & Digital twins for accurate simulation, testing and prediction of SONFs behavior & True fidelity between the physical and the digital worlds. \\ \cline{3-4}
\hline

\multirow{2}{3.5cm}{Support of backward compatibility and coexistence with legacy non-SONs or partial SONs.} & \multirow{3}{3cm}{Evolved H-SON to orchestrate network operations over various legacy and emerging technologies and standards} & Develop SON-capable gateways on the top of legacy systems with translation and normalization capabilities to endow legacy systems with automation features & Heterogeneity and complexity of legacy systems usually vendor-dependant. \\ \cline{3-4}
& & Service redundancy (i.e., develop main services on SONs and backup versions on legacy systems). & Develop seamless fallback mechanisms to downgrade the SONF to the legacy version of the service to restore an operational state in case of outage situations to improve fault tolerance. \\ \cline{3-4}
\hline

\multirow{2}{3.5cm}{The risk of losing control and situation awareness (because of emergence).} & \multirow{3}{3cm}{Evolved H-SON to guarantee a bare minimum of centralized management and control} & QoE and integration of non-network data (e.g., location-based and social-based analytics) & Information sources can have different time scales, and data structures can be complex and incompatible. High dimensionality of datasets. \\ \cline{3-4}
& & Human in the loop & Define the human intervention scope and frontiers as well as manage human weaknesses (e.g., illness) \\ \cline{3-4}
& & Advanced resilience capabilities such as self-stabilization~\cite{Dobson} to self-restore a stable system state. & The duration of the temporary system unavailability should be predictable and bounded.  \\ \cline{3-4}
\hline

\multirow{2}{3.5cm}{Fragmented self-organization capabilities.} & \multirow{2}{3cm}{Harmonized H-SON model for cellular, wired and wireless networks} & GANA as a reference model, many possible instantiations for customized implementations & New standards for the H-SON concept to define more refined degrees of centralization depending on the strength of the intended coupling. Achieve an adaptive and seamless balance between the centralized and distributed SONFs. \\ \cline{3-4}
& & SDOs/fora cooperation to unify and narrow taxonomies & A joint multi-SDO/fora development process should be adopted from the early standardization stages~\cite{IEEE_SysOpt_WG}. \\ \cline{3-4} 
\hline

\multirow{3}{3.5cm}{Computationally expensive optimizations.} & \multirow{3}{3cm}{Evolved H-SON to avoid getting stuck in local optima} & Quantum technology for faster AI and computing & Quantum capabilities in terms of running under cryogenic temperatures or very high pressures are still an open problem. \\ \cline{3-4}
& & Proactive coordination using advanced AI (e.g., FL, TL and JLC) and continuous learning & A native integration of AI capabilities with the remaining aspects of wireless networks such as computing and communication. \\ \cline{3-4}
& & Sustainable energy provisioning (e.g., large scale telemetry) & Achieving a holistic telemetry approach to avoid localized device-centric telemetry. \\ \cline{3-4}
\hline
\end{tabular}
\end{table*}
\subsection{Future Research Directions}
\label{subsec:Future_Research_Directions}
Below, we share some enabling technologies expected to lay the foundation for the evolved H-SONs.
 
\smallskip

\textbf{Virtualized, containerized, microservice, and multi-tenant architectures.} 6G will witness the full power of virtualization, containerization, microservice and multi-tenancy approaches. The use of COTS equipment from any vendor will potentially improve self-coordination capabilities through achieving additional flexibility to place a set of multi-vendor and multi-service SONFs on the top of the same virtualized environment with low cost. This can also boost network scalability, alleviate multi-vendor compatibility issues and avoid misconfiguration risks~\cite{Bonati}. On the downside, different SONFs can access the same database and consequently compromise each other’s security.
\smallskip

\textbf{Quantum technology.} Self-coordination usually needs to handle complex optimization problems, which can be resolved with quantum computing (QC). QC techniques exploit the superposition of quantum states to concurrently explore different possibilities and quickly arrive at an optimal solution. In this perspective, QC will reduce computation cost (e.g., complex calculations) and the overall latency of the network (e.g., overheads of various layers) especially in virtualized and cloudified networks. This can enhance both proactive and reactive coordination algorithms in extremely dynamic environments. However, quantum devices need to run under cryogenic temperatures or very high pressures and these conditions are still in early development stages.
\smallskip

\textbf{Federated and transfer learning.} While conventional 5G throughputs are still unable to handle the iterative transfer of large datasets over wireless links, federated learning (FL)~\cite{9210812} in 6G can simplify network automation by processing learning models (e.g., on COTS hardware installed at the edge) to reduce the time and energy costs of the proactive self-coordination. Further, transfer learning (TL) will allow capitalizing on the experience gained in avoiding or resolving previous 5G SON conflicts to address new but similar ones in 6G~\cite{Fourati}.
\smallskip
 
\textbf{Joint learning and communication.} Unlike conventional AI techniques in 5G that act fragmentally and selfishly, the ubiquitous intelligence of 6G will be integrated with various capabilities such as computing and communication into an unified design. In particular, self-coordination will leverage the power of the joint learning and communication (JLC) paradigm wherein the performance of the AI algorithms is optimized taking into account the peculiarities of the wireless channel~\cite{9210812}. For instance, inference tasks are usually split over many devices and servers. Although the intended prediction can be accurate, separately treating communication and inference may lead to violating the latency and reliability requirements of the underlying service. JLC frameworks will help mitigating such conflicts.

6G research still needs to deal with the slow convergence speed of evolutionary methods and neural networks (e.g., via offline training), and minimize the amount of data locally needed to achieve the global benefit (e.g., via FL).
\smallskip

\textbf{Contextual awareness.}
Collecting, combining, and reacting to different pieces of available contextual information, such as user location, distributed device sensors, user social activity, and related KQIs \cite{8868102}, is of paramount interest in the SON framework. The challenge for the context-aware SON is that the information sources have different time scales and they are not necessarily compatible measures having different signalling overheads and security requirements.
\smallskip

\textbf{Digital twins (DTs).} Can identify potential SON conflicts before the production phase provided that true fidelity is achieved between the DTs and their  physical counterpart.

\smallskip

\textbf{Sustainability and energy efficiency.} One way to resolve the potential conflicts is to define priorities. Coordination between SONFs attempting to optimize performance and others attempting to optimize energy usage will be an important area of future work given a goal of sustainability~\cite{Fourati}. Moreover, reliance on holistic telemetry capabilities, e.g., IETF's Network Telemetry Framework (NTF), supported by pervasive intelligence, as opposed to traditional user or device-centric mechanisms in 5G, will substantially decrease the related risks of conflicts.
\smallskip

\textbf{Human factor.} SONs aim at reducing human intervention in system optimization related tasks, but a human in the self-coordination process is indispensable to prevent system malfunction since machines are not self-conscious \cite{Bayazeed}. For a successful operational mode, human intervention should be in the highest layer loops, and ideally should not intervene in the lowest layers characterized by tighter constraints in terms of responsiveness. Moreover, and besides autonomous learning, human experts can also contribute to enriching the coordination rules fed into the network. 
\smallskip

\subsection{Concluding Discussions}
Table~\ref{tab:challenges} illustrates how the aforementioned technological breakthroughs effectively address the SON challenges identified in subsection~\ref{subsec:Challenges}. Particularly, the anticipated advances in virtualization and containerization capabilities will reduce the hardware dependability and offer the needed isolation of the SONFs with a bare minimum of adaptive centralized management. This optimal degree of centralization will be used to propagate the knowledge from the top abstraction levels down to the lowest levels to enforce the needed actions at much smaller time frames than 5G~\cite{Etsi_103_195}. The integration of a multi-tenancy dimension between various ecosystems and network parts (e.g. RAN, PCN, CCN) will boost overall SON scalability to handle billions of 6G connections. Such environments will help future H-SONs to dynamically monitor SONF instances, their mutual degrees of coupling (e.g., $C$) and their life-cycle to reduce conflicting situations.

To resolve the increasing complexity, H-SONs will exploit meta-programming to facilitate the incorporation of the quantum technology. The latter will help 6G SONs conducting expensive optimizations, and thus extending their management mechanisms with advanced AI (e.g., FL, TL and JLC). This will be supported by standardized telemetry techniques (e.g., NTF) to provide rich energy metering capabilities and bring true self-sustainability to the network.

Given the first steps towards 6G, different stakeholders should coordinate a joint SON design and development process to foresee the interworking and co-existence of Multiple Radio Access Technologies (multi-RATs) under a holistic SON standard. In turn, legacy systems can be augmented by SON-capable gateways and interfaces with translation and normalization capabilities to enforce backward compatibility.

Lastly, the QoE will be at the center of context-aware 6G SONs. New dedicated SONFs can advise about the possible actions to obtain better QoE (e.g., go to a more preferable location or change the orientation). Accordingly, emerging 6G services will be human-centric thus maintaining a significant contribution of the human factor in overall system performance and critical situations as well.

\section{Conclusion} 
\label{sec:conclusions}

This paper discussed the problem of managing complex networks more efficiently. We showed how the development of SONs has been progressing towards H-SONs. To guarantee the stability of the network, the different control feedback loops must be almost isolated from each other. This can be achieved using the concept of loose coupling. Thus we discuss the potential and the promises of loosely coupled H-SONs for an improved self-coordination to avoid potential conflicts. We also provided an in-depth landscape analysis of past and current SON-related standardization activities and the inherent self-coordination functionality within various SDOs to explore the harmonization possibilities. Finally, we summarized the challenges and the future technological trends for the development of SONs over the next decade. An important challenge is to adapt the network to the correct degree of centralization between centralized and distributed networks depending on the situation in the network environment.
\section*{Acknowledgment} 
This work was supported in part by the EU Horizon 2020 project DEDICAT 6G under Grant no. 101016499.

\bibliographystyle{IEEEtran}
\bibliography{magazine-SON-v2.bib}

\begin{thebibliography}{10}
\providecommand{\url}[1]{#1}
\csname url@samestyle\endcsname
\providecommand{\newblock}{\relax}
\providecommand{\bibinfo}[2]{#2}
\providecommand{\BIBentrySTDinterwordspacing}{\spaceskip=0pt\relax}
\providecommand{\BIBentryALTinterwordstretchfactor}{4}
\providecommand{\BIBentryALTinterwordspacing}{\spaceskip=\fontdimen2\font plus
\BIBentryALTinterwordstretchfactor\fontdimen3\font minus
  \fontdimen4\font\relax}
\providecommand{\BIBforeignlanguage}[2]{{%
\expandafter\ifx\csname l@#1\endcsname\relax
\typeout{** WARNING: IEEEtran.bst: No hyphenation pattern has been}%
\typeout{** loaded for the language `#1'. Using the pattern for}%
\typeout{** the default language instead.}%
\else
\language=\csname l@#1\endcsname
\fi
#2}}
\providecommand{\BIBdecl}{\relax}
\BIBdecl

\bibitem{Aliu}
O.~G. Aliu, A.~Imran, M.~A. Imran, and B.~Evans, ``{A survey of self
  organisation in future cellular networks},'' \emph{IEEE Communications
  Surveys \& Tutorials}, vol.~15, no.~1, pp. 336--361, First Quarter 2013.

\bibitem{Fourati}
H.~Fourati, R.~Maaloul, L.~Chaari, and M.~Jmaiel, ``{Comprehensive survey on
  self-organizing cellular network approaches applied to {5G} networks},''
  \emph{Computer Networks}, vol. 199, pp. 1--24, 9 November 2021, Art. no.
  108435.

\bibitem{Mammela}
A.~Mämmelä, J.~Riekki, A.~Kotelba, and A.~Anttonen, ``{Multidisciplinary and
  historical perspectives for developing intelligent and resource-efficient
  systems},'' \emph{IEEE Access}, vol.~6, pp. 17\,464--17\,499, 2018.

\bibitem{Dobson}
S.~Dobson, D.~Hutchison, A.~Mauthe, A.~Schaeffer-Filho, P.~Smith, and J.~P.~G.
  Sterbenz, ``Self-organization and resilience for networked systems: Design
  principles and open research issues,'' \emph{Proceedings of the IEEE}, vol.
  107, no.~4, pp. 819--834, April 2019.

\bibitem{Bayazeed}
A.~Bayazeed, K.~Khorzom, and M.~Aljnidi, ``{A survey of self-coordination in
  self-organizing network},'' \emph{Computer Networks}, vol. 196, pp. 1--32, 4
  September 2021, Art. no. 108222.

\bibitem{Kafle}
V.~P. Kafle, T.~Hirayama, T.~Miyazawa, M.~Jibiki, and H.~Harai, ``{Network
  control and management automation: Architecture standardization
  perspective},'' \emph{IEEE Communications Standards Magazine}, vol.~5, no.~3,
  pp. 106--114, September 2021.

\bibitem{mammela2022new}
A.~M{\"a}mmel{\"a} and J.~Riekki, ``{New network architectures will be weakly
  coupled},'' \emph{IEEE Future Networks Tech Focus}, April 2022.

\bibitem{IEEE_SysOpt_WG}
{IEEE International Network Generations Roadmap (INGR), Systems Optimization
  Working Group}, ``{An IEEE 5G and beyond technology roadmap: Systems
  optimization, 2021 edition},'' 2021.

\bibitem{Etsi_103_195}
{ETSI TS 103 195-2 V1.1.1}, ``{Autonomic network engineering for the
  self-managing Future Internet (AFI); Generic Autonomic Network Architecture;
  Part 2: An architectural reference model for autonomic networking, cognitive
  networking and self-management},'' May 2018.

\bibitem{Darwish}
T.~Darwish, G.~K. Kurt, H.~Yanikomeroglu, G.~Senarath, and P.~Zhu, ``{A vision
  of self-evolving network management for future intelligent vertical
  HetNet},'' \emph{IEEE Wireless Communications}, vol.~28, no.~4, pp. 96--105,
  August 2021.

\bibitem{8868102}
A.~Herrera-Garcia, S.~Fortes, E.~Baena, J.~Mendoza, C.~Baena, and R.~Barco,
  ``Modeling of key quality indicators for end-to-end network management:
  {Preparing for 5G},'' \emph{IEEE Vehicular Technology Magazine}, vol.~14,
  no.~4, pp. 76--84, 2019.

\bibitem{Long}
X.~Long, X.~Gong, X.~Que, W.~Wang, B.~Liu, S.~Jiang, and N.~Kong, ``Autonomic
  networking: Architecture design and standardization,'' \emph{IEEE Internet
  Computing}, vol.~21, no.~5, pp. 48--53, September/October 2017.

\bibitem{Martinez_Julia}
P.~Martinez-Julia, V.~P. Kafle, and H.~Harai, ``{Exploiting external events for
  resource adaptation in virtual computer and network systems},'' \emph{IEEE
  Transactions on Network and Service Management}, vol.~15, no.~2, pp.
  555--566, June 2018.

\bibitem{Bonati}
L.~Bonati, S.~D'Oro, M.~Polese, S.~Basagni, and T.~Melodia, ``{Intelligence and
  learning in {O-RAN} for data-driven {NextG} cellular networks},'' \emph{IEEE
  Communications Magazine}, vol.~59, no.~10, pp. 21--27, Oct. 2021.

\bibitem{9210812}
M.~Chen, Z.~Yang, W.~Saad, C.~Yin, H.~V. Poor, and S.~Cui, ``A joint learning
  and communications framework for federated learning over wireless networks,''
  \emph{IEEE Transactions on Wireless Communications}, vol.~20, no.~1, pp.
  269--283, 2021.

\end{thebibliography}

\begin{IEEEbiographynophoto}{Abdelaali Chaoub} [SM] has been an Associate Professor in Telecommunications, appointed at the Institut National des Postes et Télécommunications (INPT) of Morocco since 2015. He obtained an engineering degree of telecommunication from INPT in 2007 with the highest honors and received his Ph.D. degree in electrical engineering from Mohammed V-Agdal University in 2013. His research interests includes remote/rural connectivity solutions for 5G/6G networks, design and optimization of IoT-enabled smart environments, dynamic spectrum access and cognitive radio, adaptive multimedia streaming in wireless networks and mitigation of DoS attacks in cellular networks. He has worked as a Senior VoIP solutions consultant at Alcatel-Lucent (2007-2015).
\end{IEEEbiographynophoto}%

\begin{IEEEbiographynophoto}{Aarne Mämmelä} [SM] received the degree of D.Sc. (Tech.) (with honors) from the University of Oulu in 1996. He was with the University of Oulu from 1982 to 1993. In 1993 he joined the VTT Technical Research Centre of Finland in Oulu. Since 1996 he has been a Research Professor of wireless communications. He visited the University of Kaiserslautern in Germany in 1990-1991 and the University of Canterbury in New Zealand in 1996-1997. He has been an advisor to 10 doctoral students and published over 40 journal papers and over 100 conference papers. \end{IEEEbiographynophoto}%

\begin{IEEEbiographynophoto}{Pedro Martinez-Julia} [M] received his B.S. in Computer Science from the Open University of Catalonia, M.S. in Advanced Information Technology and Telematics and Ph.D. in Computer Science from the University of Murcia, Spain. Currently a full-time researcher with the National Institute of Information and Communications Technology, Tokyo. Has been involved in EU-funded research projects since 2009, leading several tasks/activities, and participating in IETF/IRTF for the new network technologies standardization. Published over twenty papers in refereed conferences/journals. His main expertise is in network architecture, control and management, with particular interest in overlay networks and distributed systems/services. Member of ACM and IEEE.\end{IEEEbiographynophoto}

\begin{IEEEbiographynophoto}{Ranganai Chaparadza} Dr.-Ing./PhD, is a Senior Capgemini Consultant for Vodafone (and other Telecommunications Network Operators) and Solutions Design Architect, Standardization of Autonomic/Autonomous Networking in ETSI TC INT/AFI WG. IPv6 Forum Fellow representative in ETSI TC INT/AFI WG. PhD in Telecomm \& Computing Engineering from Technical University of Berlin (TUB); MSc in Telecommunications Engineering from Warsaw University of Technology, Poland. Standardization Expert (ETSI, BBF, ITU-T, NGMN, TMF, 3GPP, IEEE, IETF, etc.); Innovation with SDN/NFV and AMC (Autonomic Management \& Control) using ETSI GANA-oriented standards; ETSI 5G PoC. IEEE INGR Future Networks SBB, SystOpt, and Testbeds WGs.\end{IEEEbiographynophoto}%

\begin{IEEEbiographynophoto}{Muslim Elkotob} [M] Dr.-Ing./PhD, is a Principal Solutions Architect at Vodafone with a lead role and end-to-end responsibility in the Enterprise Business Line. He works on driving innovation and standardizing architectures in SDN/NFV, Autonomics, Slicing and Security areas in 5G and IoT. An IPv6-Forum Fellow and delegate with lead roles in various SDOs including ETSI, TMForum, ITU-T and IEEE. Having a career background with vendors, service providers and R\&D, he has spent the last seven years strengthening Vodafone's role in the enterprise Value Chain as a global player with a powerful infrastructure and autonomic IT services on top.\end{IEEEbiographynophoto}%

\begin{IEEEbiographynophoto}{Lyndon Ong} [M] is Principal, Advanced Architecture in the Office of the CTO at Ciena Corporation. He is focused on control architecture and 5G/6G open RAN, and headed numerous projects on optical network control and SDN at Ciena, where he received the Ciena Technical Fellow award in 2014. Currently co-chair of the Open RAN Alliance Orchestration and Cloudification WG and Project Leader for the ONF’s Open Transport Configuration and Control Project. Previously chaired the OIF Technical Committee and the IETF Sigtran WG, where he participated in development of the SCTP protocol. He received his doctoral degree from Columbia University in 1991.\end{IEEEbiographynophoto}

\begin{IEEEbiographynophoto}{Dilip Krishnaswamy} [SM] received a PhD in electrical engineering from the University of Illinois at Urbana-Champaign. Currently serves as a Senior Principal Architect  at Sterlite Access Solutions, and recently served as VP for R\&D on new and emerging technologies at Jio Platforms. Previously he has worked as a Platform Architect at Intel,  a Senior Staff Researcher in the Office of the Chief Scientist at Qualcomm Research, and a Senior Scientist at IBM Research. Inventor on 60+ granted US patents and authored 70+ research publications. His research interests include 5G/6G networks, distributed optimization and machine learning, wireless networking, and quantum computing. \end{IEEEbiographynophoto}%

\begin{IEEEbiographynophoto}{Antti Anttonen} [SM] received the D.Sc. (Tech.) degree from the Department of Electrical and Information Engineering, University of Oulu, Finland, in 2011. He works as a Senior Scientist at VTT Technical Research Centre of Finland. He has visited Lucent Technologies, USA in 2000, University of Hannover, Germany in 2008, and University of Leuven, Belgium in 2014. He has participated in numerous national and European research projects related to wireless communications. His main interests include network management and stochastic system modelling for heterogeneous networks.
\end{IEEEbiographynophoto}

\begin{IEEEbiographynophoto}{Ashutosh Dutta} [F] is Chief 5G Strategist and Lawrence Hafstad Fellow at Johns Hopkins University Applied Physics Labs and Chair of ECE EP. Ashutosh serves as the founding Co-Chair for the IEEE Future Networks. Ashutosh served as IEEE Communications Society's Distinguished Lecturer for 2017-2020 and served as the Director of Industry Outreach for IEEE Communications Society from 2014-2019 where he currently serves as Member At Large. Ashutosh currently serves as Industry Forum and Exhibits co-chair for Globecom 2022 and the Chair for IEEE Open RAN Industry Connections. Ashutosh is a Distinguished Alumnus of NIT Rourkela, has a MS in Computer Science from NJIT, and Ph.D. in Electrical Engineering from Columbia University. Ashutosh is a Fellow of IEEE and Distinguished member ACM.
\end{IEEEbiographynophoto}

\end{document}